# Distant Jupiter family Comet P/2011 P1 (McNaught)

Pavlo P. Korsun [a,*], Oleksandra V. Ivanova [a], Viktor L. Afanasiev [b], Irina V. Kulyk [a]

[a] *Main Astronomical Observatory of National Academy of Science of Ukraine, Akademika Zabolotnoho Str. 27, 03680 Kyiv, Ukraine*
[b] *Special Astrophysical Observatory of the Russian Academy of Science, Nizhnij Arkhyz 369167, Russia*

The spectra and images obtained through broadband BVRc filters for Jupiter family Comet P/2011 P1 (McNaught) were analyzed. We observed the comet on November 24, 2011, when its heliocentric distance was 5.43 AU. Two dimensional long slit spectra and photometric images were obtained using the focal reducer SCORPIO attached to the prime focus of the 6-m telescope BTA (SAO RAS, Russia). The spectra cover the wavelength range of 4200–7000 Å. No emissions of $C_2$ and $CO^+$, which are expected in this wavelength region, were detected above $3\sigma$ level. An upper limit in gas production rate of $C_2$ is expected to be $1.1 \times 10^{24}$ mol s$^{-1}$. The continuum shows a reddening effect with the normalized gradient of reflectivity along dispersion of 5.1 ± 1.2% per 1000 Å. The color indices (B–V) = 0.89 ± 0.09 and (V–Rc) = 0.42 ± 0.07 for the nucleus region or (B–V) = 0.68 ± 0.12 and (V–Rc) = 0.39 ± 0.10 for the coma region, which are derived from the photometric data, also evidence that the color of the cometary nucleus and dust are redder with respect to the Sun. The normalized gradients of 5.9 ± 2.9% per 1000 Å and 2.6 ± 1.9% per 1000° for VRc filters were obtained for the cometary nucleus and the dust coma, respectively. The estimated dust mass production rate is about 12 kg s$^{-1}$ for Rc filter. The dust coma like a spiral galaxy edge-on was fitted using a Monte Carlo model. Since it is expected that the particles forming the dust coma consist of "dirty" ice, Greenberg's model was adopted to track grains with an icy component that evaporates slowly when exposed to solar radiation. The observed coma was fitted assuming two iso-lated active zones located at the cometocentric latitudes of −8° and −35° with outflow of the dust within the cones having half opening angles of 8° and 70°, respectively. About, 45% and 55% of the observed dust were originated from the high collimated and low collimated active zones, respectively. The spin-axis of the rotating nucleus is positioned in the comet's orbit plane. The sizes of the dust particles were ranged from 5 μm to 1 mm with a power index of −3.0 for the adopted exponential dust size distribution.

## Introduction

Comet P/2011 P1 (McNaught) (hereafter P/2011 P1) was discovered by Robert McNaught with the 0.5-m Uppsala Schmidt telescope at Siding Spring Observatory on August 1.77, 2011 (Williams, 2011). It was detected as an object of 17th magnitude that moved at distances of 5.24 and 5.17 AU from the Sun and the Earth, respectively.

The comet underwent a short-term orbit perturbation. It was clearly shown applying dynamical studies on P/2011 P1.[1] The approach to within 0.025 AU of Jupiter in 2010 December caused changes in the orbit. Prior to the latest jovian perturbations the comet was on a 12.2-year orbit with $q$ = 4.24 AU, and Incl = 4.7°. Its current orbital parameters are summarized in Table 1.[2]

Comet P/2011 P1 (McNaught) showed significant level of activity beyond the orbit of Jupiter. The coma of the comet was strongly asymmetric like a spiral galaxy edge-on. Some CCD observers commented on the object's elongated cometary appearance as a possible presence of a tail and anti-tail[3] (Green, 2011).

Our data were obtained within a program of spectroscopic and photometric investigations of distant active comets in the optical domain. The study was performed in 2006–2013 as survey of a sample of comets, which having perihelion beyond the Jupiter's orbit, showed noticeable activity. The perihelion distance of 2011P1 is somewhat closer to the Sun, nevertheless, we focused on this object due to an unusual appearance. Since there are limited data concerning the spectrophotometric investigation of Jupiter family comets (hereafter JFC) at large heliocentric distances we present here our results for a single object while extensive overviews of the photometric studies of JFC are published (Licandro et al., 2000; Lowry and Fitzsimmons, 2005; Lowry et al., 2003; Mazzotta Epifani et al., 2007, 2008; Snodgrass et al., 2008). Furthermore, we propose here a possible scenario of the observed dust coma formation using a Monte Carlo modeling.

---

[3] http://remanzacco.blogspot.com.tr/2011/08/new-comet-p2011-p1-mcnaught.html.

**Table 1**
Current orbital parameters of P/2011 P1.

| Element | Value |
|---|---|
| Eccentricity | 0.359843 |
| Semi-major axis, AU | 7.74 |
| Perihelion distance, AU | 4.96 |
| Inclination, deg | 0.0004 |
| Longitude of the ascending node, deg | 0.0017 |
| Argument of perihelion, deg | 0.0206 |
| Orbital period, yr | 21.54 |

The paper is organized as follows. The observations and reduction techniques are described in Section 2. Section 3 gives the results of analysis of the observed spectroscopic and photometric data. Model explanation of unusual appearance of the observed coma can be found in Section 4. Comparison of the derived results with similar data known for JFC and brief conclusions are presented in Section 5.

## Observations and data reduction

The observations of P/2011 P1 were made with the 6-m telescope BTA mounted at SAO RAS[4] (Special Astrophysical Observatory of Russian Academy of Sciences, Russia) on November 24, 2011, when heliocentric and geocentric distances of the comet were 5.43 and 4.54 AU, respectively. The focal reducer SCORPIO (Spectral Camera with Optical Reducer for Photometrical and Interferometrical Observations) attached to the prime focus of the telescope was operated in the photometric and long-slit spectroscopic modes (Afanasiev and Moiseev, 2005). A CCD chip E2V-42-90 with dimensions of 2K × 4K was used as a detector. The size of one pixel is 16 × 16 μm which corresponds to 0.18″ × 0.18″ on the sky plane.

The photometric data of P/2011 P1 was obtained through the B, V, and Rc broadband filters. The night was photometric and the seeing was stable around 1.5″.

Performing photometric measurements we applied a 2 × 2 binning of the original frames. The dimension of the images was 1024 × 1024 pixels and the scale was 0.36″/pix. The full field of view of the CCD is 6.1′ × 6.1′.

The reductions of the raw data, which included bias subtraction, flat field correction and cleaning from cosmic ray tracks, were made. The morning sky was exposed to provide flat field corrections for the non-uniform sensitivity of the CCD chip. The bias was removed by subtracting an averaged frame with zero exposure time, and the cosmic ray tracks were removed when we computed the averaged image from the individual ones. For this purpose, a procedure, which computes the median at each pixel across a set of 2-d images, was applied on the stacked images.

To provide the absolute photometric calibration we observed the Landolt standard stars Pg0231+051 (Landolt, 1992) and used measurements of the spectral atmospheric transparency at the Special Astronomical Observatory provided by Kartasheva and Chunakova (1978).

Spectroscopic observations were made with a long slit mask. The height of the slit was 6.1′, and the width was 1″. The transparent grism VPHG1200@540 was used as a disperser in the spectroscopic mode. The spectra cover the wavelength range of 3600–7000 Å, and the spectral resolution, which was defined by the width of the slit, is about 5 Å. The spectroscopic images were binned along the spatial direction as 2 × 1.

[4] http://www.sao.ru/Doc-en/Telescopes/.

A lamp with smoothly varying energy distribution was used to compensate for the nonuniform sensitivity of the chip's pixels in the spectroscopic mode. Wavelength calibrations were made exposing a He–Ne–Ar-filled lamp. The spectrum of the morning sky was exposed to estimate variations of the background night sky spectrum along spatial direction.

For the spectral calibration of the spectra spectrophotometric standard BD+28d4211 was observed (Oke, 1990). The spectral behavior of the atmospheric extinction was also taken from Kartasheva and Chunakova (1978).

Standard reduction manipulations regarding the obtained spectroscopic data were performed. To remove biases from the observed frames, to clean the frames from the cosmic events, and to correct their geometry we used the Scorpio_2x4K.lib package operating under IDL. The background night sky spectrum was removed using the expositions of the morning sky spectrum. The later was weighted in order to fit the observed level of the night sky spectrum in each column along the slit.

Details on the observations are collected in Table 2. Fig. 1 illustrates position of the comets during observations. Appearance of the comet, which was obtained summing the images observed through V and Rc filters, one can see in Fig. 6.

**Table 2**
Journal of observations of P/2011 P1.

| Data, UT | r, AU | Δ, AU | Exp., s | Z[a] | $P_{sun}$[b] | Data |
|---|---|---|---|---|---|---|
| Nov. 24.905, 2011 | 5.43 | 4.53 | 120 | 33 | 77.54 | Rc |
| Nov. 24.907, 2011 | 5.43 | 4.53 | 120 | 34 | 77.54 | V |
| Nov. 24.909, 2011 | 5.43 | 4.53 | 120 | 34 | 77.54 | B |
| Nov. 24.911, 2011 | 5.43 | 4.53 | 120 | 34 | 77.53 | B |
| Nov. 24.913, 2011 | 5.43 | 4.53 | 120 | 35 | 77.53 | V |
| Nov. 24.914, 2011 | 5.43 | 4.53 | 120 | 35 | 77.53 | Rc |
| Nov. 24.917, 2011 | 5.43 | 4.53 | 120 | 35 | 77.53 | Rc |
| Nov. 24.918, 2011 | 5.43 | 4.53 | 120 | 36 | 77.53 | V |
| Nov. 24.919, 2011 | 5.43 | 4.53 | 120 | 36 | 77.53 | B |
| Nov. 24.940, 2011 | 5.43 | 4.53 | 900 | 40 | 77.52 | Spectrum |
| Nov. 24.957, 2011 | 5.43 | 4.53 | 900 | 44 | 77.52 | Spectrum |
| Nov. 24.968, 2011 | 5.43 | 4.53 | 900 | 47 | 77.51 | Spectrum |

[a] Zenith distance in degrees.
[b] Position angle of the extended radius vector in degrees.

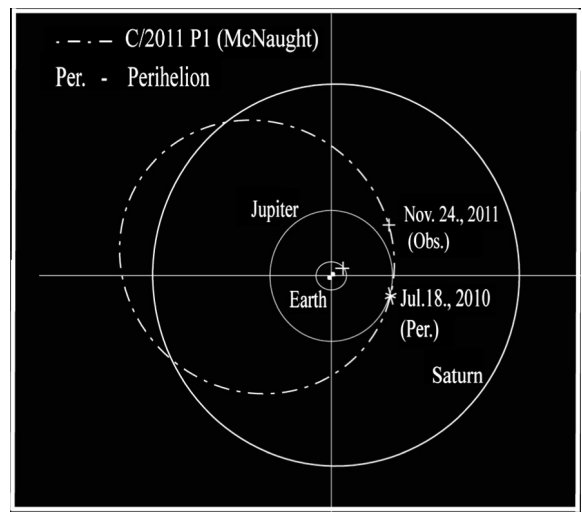

**Fig. 1.** Schematic view of the comet orbit along with the Earth's, Jupiter's, and Saturn's orbits. Position of the comet on the date of observations and the moment of the perihelion passage are marked.

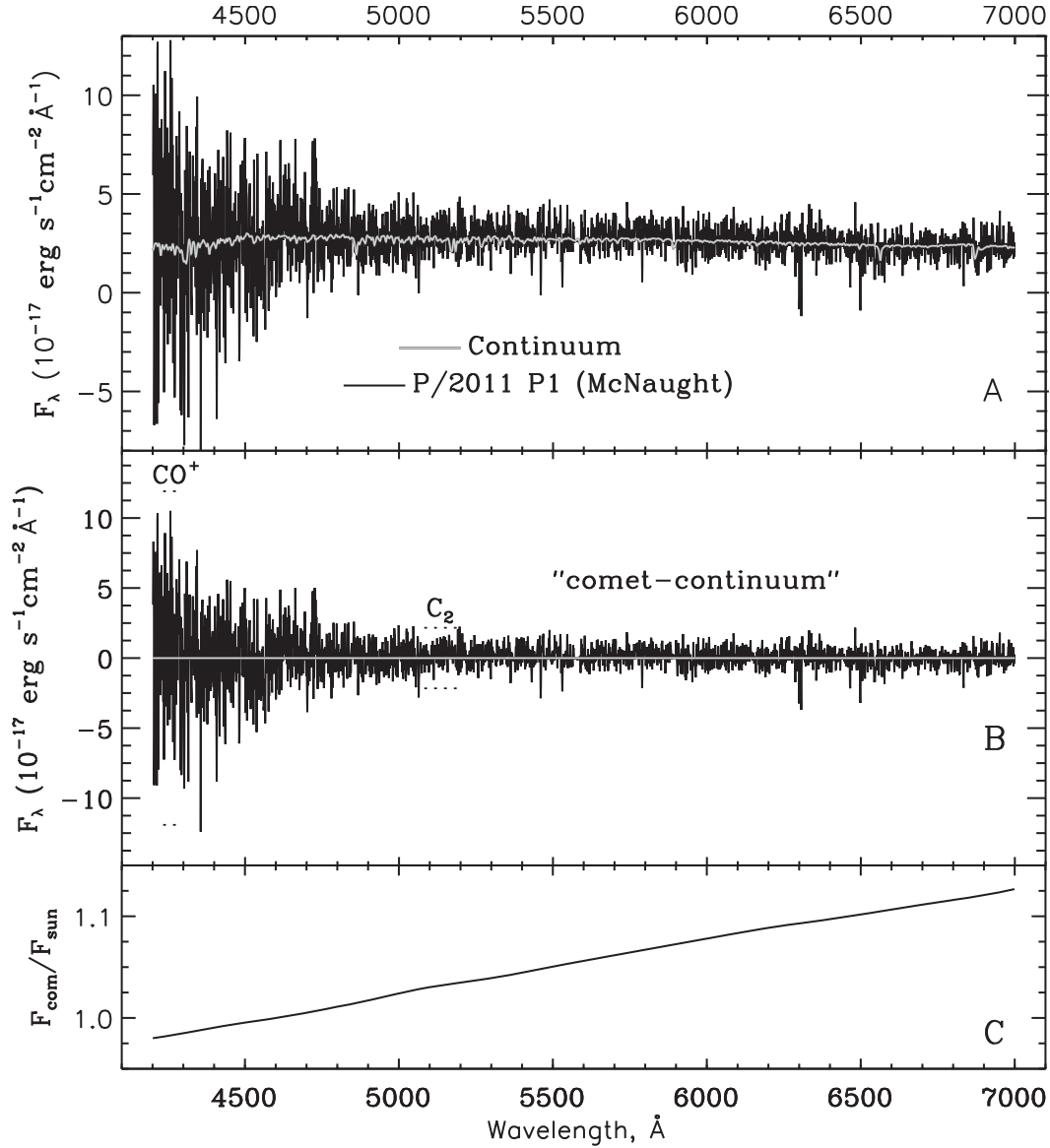

**Fig. 2.** Spectrum of P/2011 P1 with the scaled solar spectrum (A), "observed spectrum – fitted continuum" residuals, dotted lines indicate ±3σ levels, where emissions of $C_2$ and $CO^+$ are expected (B), and polynomial fitting of the (cometary spectrum)/(solar spectrum) ratio (C).

**Table 3**
Upper limits for the main cometary species not detected in the spectrum.

| Molecule | Central wavelength, Å/Δλ | Amplitude × $10^{-17}$, erg $s^{-1}$ $cm^{-2}$ $Å^{-1}$ | Flux × $10^{-17}$, erg $s^{-1}$ $cm^{-2}$ | Gas production rate, $10^{24}$ mol $s^{-1}$ |
|---|---|---|---|---|
| $CO^+$ | 4266/64 | 3.97 | <21.51 | |
| $C_2$ | 5141/118 | 0.72 | <3.90 | <1.1 |

### Analysis of observed data

S*pectra*

Three recorded two-dimensional spectra of the comet were weak and noisy. To increase the signal/noise ratio of the data to be analyzed we coadded the available spectra and collapsed the result in the spatial direction within 17 arcsec around optocenter.

Additionally, we were compelled to narrow the range of the wavelengths and cut the violet wavelengths at 4200 Å due to high level of the noise in the blue region of the spectrum. The derived behavior of the energy in the spectrum of P/2011 P1 along dispersion is displayed in Fig. 2(A). A scaled solar spectrum (Neckel and Labs, 1984) convoluted to the cometary spectrum resolution and corrected for the reddening effect is also superimposed.

To check the spectrum on the presence of possible molecular emissions we subtracted fitted continuum from the observed spectrum. The result is depicted in Fig. 2(B). Signal/noise ratio of the residuals is a strong function of wavelength. Therefore, ±3σ levels were calculated for the spectral windows, where the strongest emissions of the $C_2$ and $CO^+$ molecules are expected, and are provided as dotted lines in the figure. There are no signal above these thresholds in the above mentioned regions. So, we state that, at least, emissions of $C_2$ (the head of the strongest band (0-0) is located around 5165 Å) and $CO^+$ (the strongest band (2,0) is located within ~4250–4280 Å wavelength region) above 3σ level are not detected in the spectrum. Unfortunately, any statements on the presence of the CN and $C_3$ emissions, which strongest emissions are located at wavelengths shorter than 4200 Å, we cannot provide.

Failing to find emissions, we determined upper limits to the fluxes of $C_2$ and $CO^+$ and upper limits to the production rate of $C_2$. To obtain the upper limits to the emission fluxes, we used the same technique, which was applied to the emission-free spectrum of distant Comet C/2006 (LONEOS) (Rousselot et al., 2014). The amplitudes of the minimal measurable signal, which are equal to the RMS noise level were calculated within wavelength regions associated with the bandpass of the narrowband cometary filters (Farnham et al., 2000). The upper limits to the fluxes are listed in Table 3. An upper limit to the production rate of $C_2$ was derived using a Haser model (Haser, 1957). The model parameters were taken from A'Hearn et al. (1995). The derived upper limit to the gas production rate of $C_2$ is also listed in Table 3.

Also, we examined variation of the reflectivity $S(\lambda)$ along dispersion, which is expressed through division of the comet spectrum $F_{com}(\lambda)$ by the scaled solar spectrum $F_{sun}(\lambda)$: $S(\lambda) = F_{com}(\lambda)/F_{sun}(\lambda)$. We consider that the first degree of the polynomial fitting, which we used to fit the cometary continuum, is adequate and is depicted in Fig. 2(C). The result shows a growth of dust grain reflectivity with increasing wavelength and can be fixed quantitatively as the normalized gradient of reflectivity using

$$S'(\lambda_1, \lambda_2) = \frac{2000}{\lambda_2 - \lambda_1} \frac{S(\lambda_2) - S(\lambda_1)}{S(\lambda_2) + S(\lambda_1)}, \quad (1)$$

where $S(\lambda_2)$ and $S(\lambda_1)$ correspond to the measurements at the wavelengths $\lambda_1$ and $\lambda_2$ (in Å) with $\lambda_2 > \lambda_1$. $S'(\lambda_1, \lambda_2)$ is expressed in percent per 1000 Å. Since we adopted the first degree of the polynomial fitting, the derived reddening equal to 5.1 ± 1.1% per 1000 Å is valid for the whole examined wavelength region that is 4200–7000 Å.

The noted uncertainty is defined largely by noisiness of the analyzed spectrum. The atmospheric differential refraction affects also on the accuracy of the derived gradient. During the observations, the amplitude of the atmospheric differential refraction was about 0.5 arcsec for the treated spectral window (Filippenko, 1982). A projection of the elongation on a direction normal to the slit orientation is 0.45 arcsec as the direction of the elongation of the cometary coma is deviated from the slit orientation by 65°. Taking into account that the object is an extended source with a smooth wide surface profile we expect that an uncertainty of the gradient due to the atmospheric differential refraction is about to 0.1%. Fig. 3 provides an evidence for this conclusion.

*Photometry*

We used the images obtained through the broadband B, V and Rc filters to calculate the magnitude and dust color of P/2011 P1. The cometary magnitude is determined by

$$m_c(\lambda) = -2.5 \cdot \lg \left[ \frac{I_c(\lambda)}{I_s(\lambda)} \right] + m_{st} - 2.5 \cdot \lg P(\lambda) \cdot \Delta M, \quad (2)$$

where $m_{st}$ is the magnitude of the standard star, $I_s$ and $I_c$ are the measured fluxes of the star and the comet in counts, respectively, $P$ is the wavelength dependent sky transparency, $\Delta M$ is the difference between the comet and star airmasses.

Taking into account the seeing of the stars (1.5″) at the night of observation, we provided photometry of the data obtained through the broadband B, V and Rc filters (centered on the optocenter). We selected our apertures using the methodology presented in literature (O'Ceallaigh et al., 1995; Licandro et al., 2000; Lowry and Fitzsimmons, 2005; Mazzotta Epifani et al., 2008, 2014). We used the circular aperture with radius of 4 pixels that corresponds to 1.5″, which is equal to the seeing value measured as the average FWHM of several sample stars. The first ring is an annulus between four and nine pixels, and second ring is an annulus between nine and fourteen pixels. The $m_4(\lambda)$ value corresponds to the magnitude obtained within a circular aperture of 4 pixels, which selected flux from the nucleus with a small fraction of the coma. The $m_{4-9}(\lambda)$ magnitude was obtained within first ring, which selected the flux from the nucleus and the coma, whereas the magnitude $m_{9-14}(\lambda)$ is defined only by the cometary coma. Obtained values of the magnitudes are presented in Table 4.

We also calculated the color indices (B–V), (V–Rc) (see Table 5) and compared them with color indices of the Sun: B–V = +0.64, V–Rc = +0.35 (Holmberg et al., 2006). The dust in the coma (larger annulus) is compatible with the solar values, and that the "nucleus color" (inner circle) is comparable with values obtained for JFCs (Lamy and Toth, 2009; Snodgrass et al., 2008).

According to Jewitt and Meech (1986) the reddening can be also expressed via the normalized reflectivity gradient $S'$ (percent per $10^3$ Å):

$$S'(\lambda_1, \lambda_2) = \left( \frac{2000}{\Delta \lambda} \right) \times \frac{(10^{0.4 \cdot \Delta m} - 1)}{(10^{0.4 \cdot \Delta m} + 1)}, \quad (3)$$

where $\Delta m$ is the comet color minus the solar color. The $\Delta \lambda = \lambda_2 - \lambda_1$ parameter is the difference of the effective wavelengths of the broad-band filters pairs measured in Å. The obtained values of the normalized reflectivity gradient of P/2011 are presented in Table 5.

The detailed variation of the normalized reflectivity gradient along the detected coma was calculated using the V and Rc continuum images, which are within the wavelength interval of 5270–6470 Å. The map of the dust color is depicted in Fig. 4.

The normalized reflectivity gradient $S'$ (Fig. 4) shows variation of $S'$ within 0–6% per 1000 Å over the observed coma.

We used the obtained magnitudes to derive $Af\rho$ values. The $Af\rho$ parameter was introduced by A'Hearn et al. (1984) as a measure of the dust production rate of the cometary nucleus. The $Af\rho$ value (cm) is independent of unknown parameters, such as grain albedo or grain size. The $Af\rho$ parameter can be derived from the calculated photometric magnitude (Mazzotta Epifani et al., 2010):

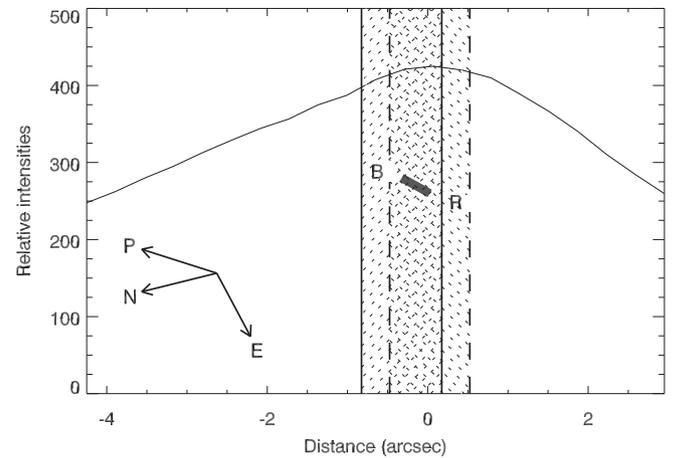

**Fig. 3.** Surface profile of the cometary coma, which was derived by summing counts in the columns of the image of the cometary coma in the direction normal to the slit direction. X axis indicates distance from the optocenter measured in arcsec. Vertical dashed strip marks bounds of the projection of the slit on the cometary coma observed through broadband Rc-filter, which corresponds to the red edge of the observed spectrum. Vertical solid strip marks bounds of the projection of the slit at 4200 Å, which corresponds to the blue edge of the observed spectrum. Short thick line represents the elongation due to the differential atmospheric refraction. Directions to North (N), East (E), and parallactic angle (P) are also provided.

**Table 4**
Photometry of P/2011 P1.

| Mag | B | V | Rc | Region shape | Aperture radius (pixel) | Aperture radius (arcsec) | Aperture radius (km) |
|---|---|---|---|---|---|---|---|
| $m_4$ | 22.07 ± 0.07 | 21.18 ± 0.06 | 20.76 ± 0.04 | Circle | 4 | 1.5 | 4895 |
| $m_{4-9}$ | 21.64 ± 0.09 | 20.92 ± 0.08 | 20.55 ± 0.06 | Annulus | 4–9 | 1.5–3.2 | 4895–10,443 |
| $m_{9-14}$ | 22.43 ± 0.09 | 21.75 ± 0.08 | 21.36 ± 0.06 | Annulus | 9–14 | 3.2–5.04 | 10,443–16,318 |

**Table 5**
Color indices and normalized reflectivity gradient of P/2011 P1 for selected photometric regions.

| Mag | B–V | V–Rc | $S'$, % | |
|---|---|---|---|---|
| | | | BV | VRc |
| $m_4$ | 0.89 ± 0.09 | 0.42 ± 0.07 | 15.6 ± 7.1 | 5.9 ± 2.9 |
| $m_{4,9}$ | 0.72 ± 0.12 | 0.37 ± 0.10 | 6.3 ± 2.4 | 2.1 ± 1.7 |
| $m_{9,14}$ | 0.68 ± 0.12 | 0.39 ± 0.10 | 3.1 ± 2.1 | 2.6 ± 1.9 |

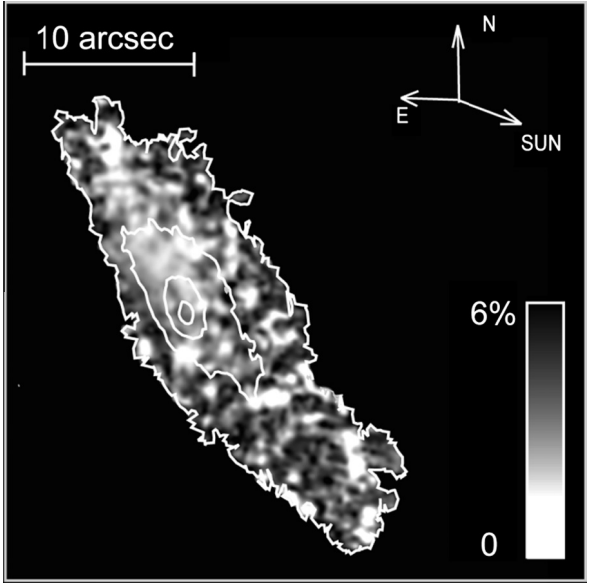

**Fig. 4.** Smoothed (over 1.5 arcsec) map of the normalized reflectivity gradient derived from the V and Rc images of P/2011 P1. Isophotes of the distribution of brightness along the coma, which is the sum of V and Rc images, are also superimposed. Directions to North, East, the Sun, and a color bar are provided.

$$Af\rho = \frac{4 \cdot r^2 \cdot \Delta^2 \cdot 10^{0.4 \cdot (m_{SUN} - m_c)}}{\rho}, \quad (4)$$

where $A$ is Bond albedo, $f$ is the filling factor in the aperture field of view, and $\rho$ is the linear radius of the aperture at the comet, i.e. the sky-plane radius, $m_{SUN}$ is the magnitude of the Sun, $r$ in AU and $\Delta$ in cm are the heliocentric and geocentric distances, respectively.

We estimated the $Af\rho$ value for the Comet P/2011 P1 for B, V, and Rc filters, taking into account that no emissions were detected in the cometary spectrum. For the calculation we used the aperture radius $\rho = 5.04''$, which is corresponded to 16,318 km. The $Af\rho$ value is 119 ± 8 cm for B filter, 128 ± 7 cm for V filter, and 114 ± 4 cm for Rc filter.

Using the obtained $Af\rho$ values, we estimated the dust mass production rate of the comet (Newburn and Spinrad, 1985; Weiler et al., 2003). The relation to estimate the dust mass rate is given by

$$Q_M = Q_N(4\pi/3) \cdot \left[\int_{a_{min}}^{a_{max}} \rho_d(a) \cdot a^3 \cdot f(a)da\right], \quad (5)$$

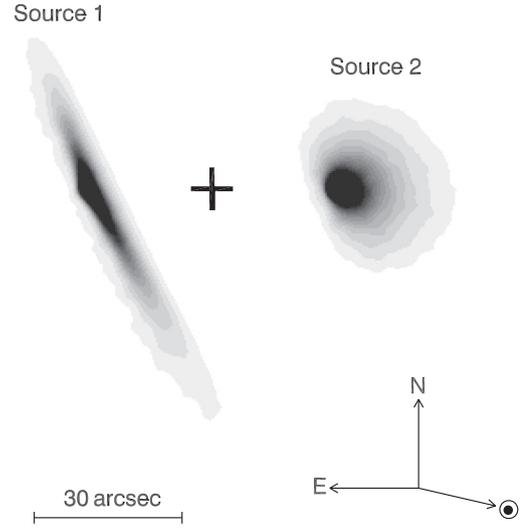

**Fig. 5.** Contributions of active areas (Source 1, left panel; Source 2, right panel) to the modeled coma. The scale bar of 30″, directions to North, East, and the Sun are also marked.

where $\rho_d(a) = \rho_0 - \rho_1(a/(a + a_2))$ is the grain density (Newburn and Spinrad, 1985), with $\rho_0 = 3000$ kg m$^{-3}$, $\rho_1 = 2200$ kg m$^{-3}$, and $a_2 = 2$ μm; $f(a)$ is the differential particle size distribution, where $a$ is the grain radius; $a_{min}$ and $a_{max}$ are the minimal and maximal grain radii. We used the lower and upper limits of dust grain radii equal to 5 and 1000 μm, respectively, based on the results obtained from the numerical modeling of dust environments of the comet. The dust number rate is given by

$$Q_N = Af\rho \cdot [2 \cdot \pi^2 \cdot p(\lambda) \cdot \Phi(\alpha)]^{-1} \cdot \left[\int_{a_{min}}^{a_{max}} (f(a) \cdot a^2 / v(a))da\right]^{-1}. \quad (6)$$

At the moment of observations the comet phase angle was $\alpha = 4.6°$. We adopted the value of geometric albedo equal to $p(\lambda) = 0.05$ for all wavelengths (Hanner and Newburn, 1989; Kolokolova et al., 2004) and the phase darkening to $\Phi(\alpha) = 0.8255$.[5]

The dust outflow velocity $v(a)$ was calculated with the equation taken from Sekanina et al. (1992), and we used velocities from 4 m s$^{-1}$ (for 1000 μm) to 52 m s$^{-1}$ (for 5 μm). The grain size limits and the dust outflow velocity will be discussed below in Section 3.3. To calculate the dust mass production rates we had to model the particle size distribution in the simple form of $f(a) \sim a^{-n}$, were $n = 3$ was obtained from the modeling of the dust coma of the comet (Section 3.3). The equation for dust density was taken from Newburn and Spinrad (1985). The dust mass production rates $Q_d$ is 9.7 kg/s for B filter, 10.6 kg/s for V filter and 12.1 kg/s for Rc filter, respectively.

*Monte Carlo modeling*

To fit the coma we used a Monte Carlo model that was described in detail by Korsun et al. (2010). Since we have only one-epoch observation of the comet, we suppose that we present just one of the solutions that could explain the observed dust environment.

---
[5] http://asteroid.lowell.edu/comet/dustphaseHM_table.txt.

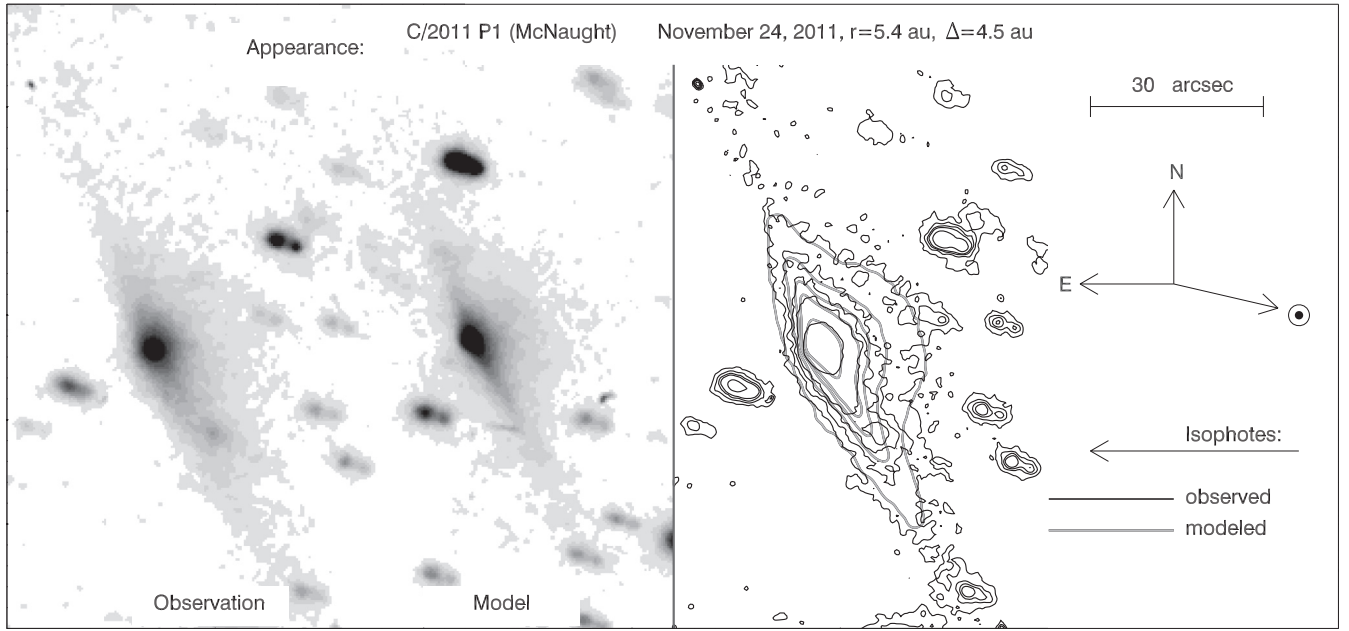

**Fig. 6.** Dust environment of P/2011 P1 observed through V and Rc filters. The modeled brightness distribution is overlaid on the observed images and shifted for clarity (left panel). The right panel illustrates the contour plots of observed and simulated brightness with dark and gray lines, respectively. The scale bar of 30″, directions to North, East, and the Sun are marked.

P/2011 P1 was observed at a heliocentric distance of 5.43 AU, where sublimation of water ice, the major volatile in cometary nuclei, is inefficient. Consequently, it seems reasonable to expect that the particles forming the dust coma consist of "dirty" ice and Greenberg's model was adopted to follow an icy grain component that evaporates slowly when exposed to solar radiation (Greenberg and Hage, 1990). The presence of water–ice grains in periodic comets is now well established during space encounters (A'Hearn et al., 2011; Schulz et al., 2006).

When the comet was discovered, on August 1, 2011, a coma surrounding the nucleus was appeared. Therefore activity of the nucleus was started earlier, and beginning of 2011 was fixed as starting time in our model runs.

A problem was to adopt a scenario of formation of the dust coma seen as a spiral galaxy edge-on. An idea, that the observed coma was formed due to activity of isolated areas located on the surface of the rotating nucleus of the comet was suggested. The domination of jet-like activity was recently well established for Comet 67P/Churyumov–Gerasimenko with the scientific imaging system OSIRIS[6] (Sierks et al., 2015).

We started our modeling assuming a clockwise rotating nucleus with one isolated active zone (Source 1). An encouraging result in conformity with orientation and extension of the observed coma, which is depicted in Fig. 5 (left panel), was obtained assuming the spin-axis of the rotating nucleus was highly deviated from the normal to the comet's orbit plane. It should be noted that the considerable obliquity of the rotational axis of Comet 67P/Churyumov–Gerasimenko equal to 52° was also determined by Sierks et al. (2015). The best fitting was achieved, when the spin axis of the nucleus was deviated from the normal by 90°, that is the spin axis was located in the comet's orbit plane. The active area (Source 1) was located to the south from the cometary equator at the cometocentric latitude of $-8°$. Outflow of the dust from this area was within a jet and was formalized as a cone shaped emitting of the solid particles. The value of the half opening angle of the cone, which satisfied the observed image, was found to be about 8°.

[6] https://rosetta.jpl.nasa.gov/news/fine-structure-activity-jets-67p/c-g.

The left panel of Fig. 5 shows the contribution of the active area (Source 1) to the modeled coma. It is obvious (see Fig. 6, below) that the one active area cannot reproduce adequately the observed coma. Consequently, we fixed the derived spin-axis of the nucleus and the position of the active area and examined more sources of the dust flowing into the cometary coma. Additional ejecta also located to the south from the cometary equator at the cometocentric latitude of $-35°$ resolved the discrepancy (Fig. 5, right panel). This outflow was substantially uncollimated and we used a cone approximation with a half opening angle of 70° to define it. The derived large opening angle implies that the dust emission originates from either one large active area or from a group of isolated minor spots.

Comparable amounts of dust were ejected toward the coma of the comet from both active zones, namely, 45% and 55% for the collimated and uncollimated outflow, respectively. The ratio of the amounts of dust was derived fitting the surface brightness distribution in the coma of the comet.

Escaping velocities of the dust, $v_d$, into the narrow cone with respect to uncollimated outflow were higher by a factor 2.5. We determined them using a relation proposed by Sekanina et al. (1992):

$$v_d = (A_0 + B_0\sqrt{a})^{-1} r^{-0.5}, \qquad (7)$$

where $A_0$ and $B_0$ are model parameters, factor $r^{-0.5}$ is heliocentric dependence of the velocities. The values of the $A_0$ and $B_0$ parameters, that we used in the model runs and specific values for the velocity of particles with 5 μm dimension are provided in Table 6.

**Table 6**
Model parameters used for our dust environment modeling.

| Parameter | Source 1 | Source 2 |
| --- | --- | --- |
| Start activity | January 1, 2011 | January 1, 2011 |
| Inclination of the spin axis (°) | 90 | 90 |
| Source latitude (°) | −8 | −35 |
| Collimation | Cone, half angle = 8° | Cone, half angle = 70° |
| Relative dust production (%) | 45 | 55 |
| Solid sizes (μm) | 5–1000 | 5–1000 |
| Size distribution index $a^x$ | −3.0 | −3.0 |
| $B_0$ (Eq. (7)), $A_0 = 0$ | 0.0053 | 0.0021 |
| Velocities ($a = 5$ μm, $r = 5.43$ AU) (m s$^{-1}$) | 36.2 | 90.6 |

The amount of the dust emitting from both active zones was found to be sunward sensitive. To fit the observed coma the dust particles were only emitted into the dayside hemisphere of the nucleus.

Wide size range of the dust particles from 5 μm to 1 mm was examined in the modeling. An exponential dust size distribution was adopted and the best result was gained with power index of −3.0.

A trial and error approach was used to fit the observed coma. Minimization of the sum of the squares of the deviations of modeled intensities from the observed ones was chosen as the quality of the fit. The final result of our modeling efforts is displayed in Fig. 6. We put the modeled image into the observed frame to get the same noisy background as for the observed one. There is comparison of the observed and modeled isophotes in the right panel of the figure and we consider the result is well acceptable. The parameters we used to fit the coma are listed in Table 6.

**Discussion and conclusions**

P/2011 P1 is the Jupiter family comet, which orbit was perturbed by Jupiter's gravitation about one year before the observation date. Active appearance of the comet was detected in August 2011, that is about half of the year after close encounter with Jupiter. It is not excluded that appearance of the conspicuous comet activity is associated with a change in its orbital parameters. High degree of correlation of these events for the Jupiter-family comets at large heliocentric distances was derived by Licandro et al. (2000). Such correlation was not confirmed by Mazzotta Epifani et al. (2007, 2008) in their investigation of the distant activity of short period comets. Moreover, they emphasize that the effect would be occurred due to possible increase of the blackbody temperature of the nucleus in case of the inward 'jump' in the perihelion distance. On the contrary, the perihelion distance of P/2011 P1 was increased from 4.24 AU to 4.86 AU. Consequently, if a connection between these events still occurs, the cause of the appearance of the activity of the comet may be local destruction of the dust crust on the surface of the comet's nucleus due to the gravitational influence of Jupiter, as their encounter in 2010 December was rather close, only 0.025 AU. If so, then the encounter justifies the adopted start of the activity in our model runs.

The level of the cometary activity (the value $Af\rho$ is 114 ± 4 cm for Rc filter), despite a heliocentric distance of 5.43 AU, was rather high comparing with other members of the population. Only a few JFC, such as 159P/LONEOS, P/2004 V5 (LINEAR-Hill), and P/2002 T5 (LINEAR), show comparable or higher level of activity at large heliocentric distances, while most of them demonstrate low level activity or are even inactive (Licandro et al., 2000; Lowry and Fitzsimmons, 2005; Lowry et al., 2003; Mazzotta Epifani et al., 2007, 2008; Snodgrass et al., 2008).

The mean value of the normalized gradient derived from our spectroscopic data is 5.1 ± 1.2% per 1000 Å. The normalized gradients of 5.9 ± 2.9% per 1000 Å and 2.6 ± 1.9% per 1000 Å for VRc filters were obtained for the cometary nucleus and the dust coma, respectively. The results, being somewhat different as the related wavelength regions and the examining portions of the coma are not the same, yet agree with each other within the error bars. Nevertheless, we fix a trend toward bluer color at larger radii. Such tendency was noted for a number of active Centaurs (Jewitt, 2009) and, for example, for Comet 10P/Tempel 2 (A'Hearn et al., 1989) or for Comet 67P/Churyumov–Gerasimenko (Snodgrass et al., 2013).

The reddening can be also presented as a comparison of values of color indices of the comet and the Sun. The dust coma has color index of V–Rc = 0.39 ± 0.10. Many of the known colors (or normalized gradients) measured so far for the JFC are redder than the Sun showing a wide diversity in values (Hainaut et al., 2012; Lamy and Toth, 2009). For a limited number of JFC being active at large heliocentric distances (>4 AU) the colors of the nuclei and their dust environment were measured (Lacerda, 2013; Lowry et al., 2003; Snodgrass et al., 2008). The slightly red color (V–R) = 0.47 ± 0.06 mag was determined for the nucleus region of Comet P/2010 TO20 LINEAR-Grauer (Lacerda, 2013) and the noticeably blue value of 0.16 ± 0.09 mag measured for Comet 103P/Hartley 2, the target for the Deep Impact extended mission EPOXI (Snodgrass et al., 2008). Our values of the reflectivity gradients for Comet P/2011 P1 (McNaught) fall in the 0–13%/1000 Å range that are typical for JFC (Snodgrass et al., 2008).

Short period comets probably originate from Centaurs, therefore there is a reason to compare the derived colors with those of active Centaurs. Most of known active Centaurs, except for P/2001 T4 (NEAT), show the neutral/slightly red color, which is correlated with our data (Bauer et al., 2003; Hartmann et al., 1990; Jewitt, 2009; Mazzotta Epifani et al., 2011, 2014).

The estimated dust mass production rate is about 12 kg s$^{-1}$ for the Rc filter, using value of the mean geometric albedo 0.05 and the grain outflow velocities from 4 m s$^{-1}$ (for 1000 μm) to 52 m s$^{-1}$ (for 5 μm). A few data on dust mass production rates derived for JFC are known. A wide range of dust mass production rates, 10–900 kg s$^{-1}$, which is higher than our result, is provided for well-known large Comet 29P/Schwassmann–Wachmann 1 in the publications (Jewitt, 1990; Moreno, 2009). Nevertheless, our result agrees with the estimations of dust mass production rates obtained for Comets 46P/Wirtanen and 67P/Churyumov–Gerasimenko during their 2002/2003 apparition at the distances closer to the Sun (Kidger, 2004, Fig. 4) or with corridor of 0.1–15 kg s$^{-1}$ derived for Centaur-comet "transition" object P/2010 C1 (Scotti), which was observed at a comparable heliocentric distance (Mazzotta Epifani et al., 2014). The dust loss rate of 7 kg s$^{-1}$ at heliocentric distances from 3.6 to 3.4 AU was derived by interpolation between the GIADA and OSIRIS data for Comet 67P/Churyumov–Gerasimenko (Rotundi et al., 2015).

We didn't detect any emissions in the 4200–7000 Å spectral window. An upper limit in gas production rate of $C_2$ is expected to be $1.1 \times 10^{24}$ mol s$^{-1}$. The upper limit, which was estimated for Rosetta target Comet 67P/Churyumov–Gerasimenko at heliocentric distance of 3.22 AU is $3.1 \times 10^{23}$ mol s$^{-1}$ (Schulz et al., 2004) and $2.6 \times 10^{23}$ mol s$^{-1}$ at heliocentric distance of 2.81 AU for 46P/Wirtanen (Schulz et al., 1998).

We fitted the dust coma of P/2011 P1 which was appeared like a spiral galaxy edge-on using a Monte Carlo model. Since it is expected that the particles forming the dust coma consist of "dirty" ice, Greenberg's model was adopted to describe the chemical composition of the particles. The observed coma was fitted assuming two isolated active zones located at the cometocentric latitudes of −8° and −35° with outflow of the dust within the cones having half opening angles of 8° and 70°, respectively. The non-isotropic emission from comets has been known for decades. Evidence of dust jets in Comet Halley was provided from ground-based and spacecraft imaging (Sekanina and Larson, 1986). Recent high-resolution spacecraft observations exhibit isolated jet features emanating from the comet's surface, indicating that comets do not emit isotopically, but instead have regions of concentrated productivity and other regions that have little or no activity (Farnham et al., 2013; Keller et al., 1987; Sekanina et al., 2004; Sierks et al., 2015; Soderblom et al., 2004; Syal et al., 2013). About 45% and 55% of the observed dust were originated from the active zones locating at the cometocentric latitudes of −8° and −35°, respectively. The spin-axis of the rotating nucleus is positioned in the comet's orbit plane. The sizes of the dust particles were ranged from 5 μm to 1 mm with power index of −3.0 for adopted exponential dust size distribution.

## Acknowledgments

The observations, which were made within the framework of a program of spectroscopic and photometric investigations of distant active comets in the optical domain at the 6-m telescope BTA, were performed due to the support of the Schedule Committee for Large Telescopes (Russian Federation). The authors thank the anonymous reviewers for valuable comments on the manuscript.

## References

A'Hearn, M. et al., 1984. Comet Bowell 1980b. Astron. J. 89, 579–591.
A'Hearn, M.F. et al., 1989. The nucleus of Comet P/Tempel 2. Astron. J. 347, 1155–1166.
A'Hearn, M.F. et al., 1995. The ensemble properties of comets: Results from narrowband photometry of 85 comets, 1976–1992. Icarus 118, 223–270.
A'Hearn, M.F. et al., 2011. EPOXI at Comet Hartley 2. Science 332, 1396–1400.
Afanasiev, V.L., Moiseev, A.V., 2005. The SCORPIO universal focal reducer of the 6-m telescope. Astron. Lett. 31, 194–204.
Bauer, J.M., Fernandez, Y.R., Meech, K.J., 2003. An optical survey of the active Centaur C/NEAT (2001 T4). Publ. Astron. Soc. Pacific 115, 981–989.
Farnham, T.L., Schleicher, D.G., A'Hearn, M.F., 2000. The HB narrowband comet filters: Standard stars and calibrations. Icarus 147, 180–204.
Farnham, T.L. et al., 2013. Connections between the jet activity and surface features on Comet 9P/Tempel 1. Icarus 222, 540–549.
Filippenko, A.V., 1982. The importance of the atmospheric differential refraction in the spectrophotometry. Publ. Astron. Soc. Pacific 94, 715–721.
Green, D.W.E., 2011. Comet P/2011 P1 (McNaught). CBET 2779.
Greenberg, J.M., Hage, J.I., 1990. From interstellar dust to comets: A unification of observational constraints. Astrophys. J. 361, 260–274.
Hainaut, O.R., Boehnhardt, H., Protopapa, S., 2012. Colours of minor bodies in the outer Solar System. II. A statistical analysis revisited. Astron. Astrophys. 546, 20pp A115.
Hanner, M.S., Newburn, R.L., 1989. Infrared photometry of Comet Wilson (1986) at two epochs. Astrophys. J. 97, 254–261.
Hartmann, W.K. et al., 1990. 2060 Chiron: Colorimetry and cometary behavior. Icarus 83, 1–15.
Haser, L., 1957. Distribution d'intensite dans la tete d'une cometes. Bull. Soc. R. Sci. Liege 43, 740–750.
Holmberg, J., Flynn, C., Portinari, L., 2006. The colours of the Sun. Mon. Not. R. Astron. Soc. 367 (2), 449–453.
Jewitt, D., 1990. The persistent coma of Comet P/Schwassmann–Wachmann 1. Astrophys. J. 351, 277–286.
Jewitt, D.C., 2009. The active Centaurs. Astron. J. 137, 4296–4312.
Jewitt, D., Meech, K., 1986. Cometary grain scattering versus wavelength, or what color is comet dust? Astrophys. J. 310, 937–952.
Kartasheva, T.A., Chunakova, N.M., 1978. Spectral atmospheric transparency at the Special Astrophysical Observatory of the USSR Academy of Science from 1974 to 1976. Astrof. Issled. Izv. Spets. Astr. Obs. 10, 44–51 (Rus.).
Keller, H.U. et al., 1987. Comet P/Halley's nucleus and its activity. Astron. Astrophys. 187, 807–823.
Kidger, M.R., 2004. Dust production and coma morphology of 67P/Churyumov–Gerasimenko during the 2002/2003 apparition. II. A comparative study of dust production in 46P/Wirtanen and 67P/Churyumov–Gerasimenko during their 2002/2003 apparition. Astron. Astrophys. 420, 389–395.
Kolokolova, L. et al., 2004. Physical properties of cometary dust from light scattering and thermal emission. In: Festou, M.C., Keller, H.U., Weaver, H.A. (Eds.), Comets II. University of Arizona Press, Tucson, pp. 577–604.
Korsun, P.P. et al., 2010. Dust tail of the active distant Comet C/2003 WT42 (LINEAR) studied with photometric and spectroscopic observations. Icarus 210, 916–929.
Lacerda, P., 2013. Comet P/2010 TO20 LINEAR-Grauer as a mini-29P/SW1. Mon. Not. R. Astron. Soc. 428, 1818–1826.
Lamy, P., Toth, I., 2009. The colors of cometary nuclei—Comparison with other primitive bodies of the Solar System and implications for their origin. Icarus 201, 674–713.
Landolt, A.U., 1992. UBVRI photometric standard stars in the magnitude range 11.5–16.0 around the celestial equator. Astron. J. 104 (1), 340–371, 436–491.
Licandro, J. et al., 2000. CCD photometry of cometary nuclei, I: Observations from 1990–1995. Icarus 147, 161–179.
Lowry, S.C., Fitzsimmons, A., 2005. William Herschel Telescope observations of distant comets. Mon. Not. R. Astron. Soc. 358, 641–650.
Lowry, S.C., Fitzsimmons, A., Collander-Brown, S., 2003. CCD photometry of distant comets. III. Ensemble properties of Jupiter-family comets. Astron. Astrophys. 397, 329–343.
Mazzotta Epifani, E. et al., 2007. The distant activity of short-period comets – I. Mon. Not. R. Astron. Soc. 381, 713–722.
Mazzotta Epifani, E. et al., 2008. The distant activity of short-period comets – II. Mon. Not. R. Astron. Soc. 390, 265–280.
Mazzotta Epifani, E. et al., 2010. The activity of Comet C/2007 D1 (LINEAR) at 9.7 au from the Sun. Astron. Astrophys. 513, 513–517.
Mazzotta Epifani, E. et al., 2011. The cometary activity of Centaur P/2004 A1 (LONEOS). Mon. Not. R. Astron. Soc. 415, 3097–3106.
Mazzotta Epifani, E. et al., 2014. Blending the distinctions among groups of minor bodies: A portrait of the Centaur-comet "transition" object P/2010 C1 (Scotti). Astron. Astrophys. 565, 8pp A69.
Moreno, F., 2009. The dust environment of Comet 29P/Schwassmann–Wachmann 1 from dust tail modeling of 2004 near-perihelion observations. Astrophys. J. Suppl. 183, 33–45.
Neckel, H., Labs, D., 1984. The solar radiation between 3300 and 12,500 Å. Sol. Phys. 90, 205–258.
Newburn, R.L., Spinrad, H., 1985. Spectrophotometry of seventeen comets. II – The continuum. Astron. J. 90, 2591–2608.
O'Ceallaigh, D.P., Fitzsimmons, A., Williams, I.P., 1995. CCD photometry of Comet 109P/Swift-Tuttle. Astron. Astrophys. 297, L17–L20.
Oke, J.B., 1990. Faint spectrophotometric standard stars. Astron. J. 99, 1621–1631.
Rotundi, A. et al., 2015. Dust measurements in the coma of Comet 67P/Churyumov–Gerasimenko inbound to the Sun. Science 347, 6pp aaa3905.
Rousselot, P. et al., 2014. Monitoring of the cometary activity of distant Comet C/2006 S3 (LONEOS). Astron. Astrophys. 480, 9pp A73.
Schulz, R. et al., 1998. Spectral evolution of Rosetta target Comet 46P/Wirtanen. Astron. Astrophys. 335, L46–L49.
Schulz, R., Stüwe, J.A., Boehnhardt, H., 2004. Rosetta target Comet 67P/Churyumov–Gerasimenko. Postperihelion gas and dust production rates. Astron. Astrophys. 422, L19–L21.
Schulz, R. et al., 2006. Detection of water ice grains after the Deep Impact onto Comet 9P/Temple 1. Astron. Astrophys. 448, L53–L56.
Sekanina, Z., Larson, S.M., 1986. Dust jets in Comet Halley observed by Giotto and from the ground. Nature 321, 357–361.
Sekanina, Z. et al., 1992. Major outburst of periodic Comet Halley at a heliocentric distance of 14 au. Astron. Astrophys. 263, 367–386.
Sekanina, Z. et al., 2004. Modeling the nucleus and jets of Comet 81P/Wild 2 based on the Stardust encounter data. Science 304, 1769–1774.
Sierks, H. et al., 2015. On the nucleus structure and activity of Comet 67P/Churyumov–Gerasimenko. Science 347, 5pp aaa1044.
Snodgrass, C., Lowry, S.C., Fitzsimmons, A., 2008. Optical observations of 23 distant Jupiter family comets, including 36P/Whipple at multiple phase angles. Mon. Not. R. Astron. Soc. 385, 737–756.
Snodgrass, C. et al., 2013. Beginning of activity in 67P/Churyumov–Gerasimenko and predictions for 2014–2015. Astron. Astrophys. 557, 15pp A33.
Soderblom, L.A. et al., 2004. Imaging Borrelly. Icarus 167, 4–15.
Syal, M.B. et al., 2013. Geologic control of jet formation on Comet 103P/Hartley 2. Icarus 222, 610–624.
Weiler, M. et al., 2003. The dust activity of Comet C/1995 O1 (Hale–Bopp) between 3 AU and 13 AU from the Sun. Astron. Astrophys. 403, 313–322.
Williams, G.V., 2011. Comet P/2011 P1 (McNaught). MPEC 2011-P19.